\documentclass[aps,prl,reprint,superscriptaddress,amsmath,amsfonts,amssymb]{revtex4-2}
\usepackage[T1]{fontenc}
\usepackage[spanish,USenglish]{babel}
\usepackage[colorlinks,allcolors=blue]{hyperref}
\usepackage{braket}
\usepackage{bm}
\usepackage{microtype}
\usepackage{graphicx}
\DeclareMathAlphabet{\mathbbold}{U}{bbold}{m}{n}

\begin{document}

\title{Is \texorpdfstring{$\chi_{c1}(3872)$}{X(3872)} generated from string breaking?}
\author{R. Bruschini}
\email{roberto.bruschini@ific.uv.es}
\affiliation{\foreignlanguage{spanish}{Unidad Te\'orica, Instituto de F\'isica Corpuscular (Universidad de Valencia--CSIC), E-46980 Paterna (Valencia)}, Spain}
\author{P. Gonz\'alez}
\email{pedro.gonzalez@uv.es}
\affiliation{\foreignlanguage{spanish}{Unidad Te\'orica, Instituto de F\'isica Corpuscular (Universidad de Valencia--CSIC), E-46980 Paterna (Valencia)}, Spain}
\affiliation{\foreignlanguage{spanish}{Departamento de F\'isica Te\'orica, Universidad de Valencia, E-46100 Burjassot (Valencia)}, Spain}

\begin{abstract}
We show, from a diabatic analysis of Lattice results for string breaking, that  mixing of $Q\bar{Q}$ with open-flavor meson-meson configurations may be expressed through a mixing potential which is order $1 / m_{Q}$. A relation between the minimum string breaking energy gap and the string tension comes out naturally. Using this relation, and matching the energy gap for $b\bar{b}$ with Lattice QCD data, we study the mixing in the $c\bar{c}$ case without any additional parameter. A $1^{++}$ bound state very close below the ${D}^{0}\bar{D}^{\ast 0}$  threshold, in perfect correspondence with $\chi_{c1}(3872)$, is predicted.
\end{abstract}

\maketitle

A new era in heavy-quark meson spectroscopy begun in 2003 with the discovery by the Belle collaboration of the $X(3872)$ \cite{Cho03}, now labeled $\chi_{c1}(3872)$ by the Particle Data Group (PDG) \cite{PDG20}, a $J^{PC}=1^{++}$ meson containing a $c\bar{c}$ component, but with properties at odds with those expected for a conventional charmonium state. Since then, many other unconventional quarkoniumlike mesons have been discovered, the relevant role played by open-flavor meson-meson components has been recognized, and an enormous theoretical effort has been dedicated to their description, see, for instance, \cite{Bra11,Bod13,Chen16,Hosaka16,Dong17,Esp17,Leb17,Guo18,Olsen18,Liu19,Yuan19, Bra20} for a comprehensive review. Notwithstanding this, almost twenty years after the discovery of $\chi_{c1}(3872)$, a deep dynamical understanding of unconventional heavy-quark mesons is still lacking. In this letter, we aim at contributing to this understanding through the analysis of the dynamical role played by string breaking. We show from Lattice results that string breaking may be the essential dynamical ingredient for a QCD-based explanation of the new states, in particular $\chi_{c1}(3872)$ on which we focus.

For this purpose, we use the recently developed Diabatic Approach in QCD, which allows for a unified and consistent description of conventional and unconventional heavy-quark mesons from Lattice inputs \cite{Bru20,Bru21c}. To be more precise, let us consider a heavy-quark meson characterized by quantum numbers $J^{PC}$, made of a heavy quark-antiquark $Q\bar{Q}$ (with $Q$ a heavy quark, $b$ or $c$) component and two open-flavor meson-meson components, $M^{(i)}\bar{M}^{(i)}$ with $i=1,2$, where $\bar{M}^{(i)}$ does not indicate necessarily the antiparticle of $M^{(i)}$, interacting with the light (gluon and sea quark) fields. From this case, the generalization to a different number of meson-meson components is straightforward.

Notice that the mass $m_{M}$ of any open-flavor meson can be written as $m_{M}=m_{Q}+c$, where $m_{Q}$ is the heavy quark mass and $c$ includes the light quark mass and the binding energy. Hence, for ${c}/ m_{Q}\ll1$, as it will be our case, the center of mass of $M^{(i)}\bar{M}^{(i)}$ practically coincides with that of $Q\bar{Q}$. On the other hand, given the large ratio of the heavy-quark mass, $m_{Q}$, to the QCD energy scale associated to the light fields, $\Lambda_{QCD}$, the instantaneous configuration of these fields can be approximately determined by considering static color sources $Q$ and $\bar{Q}$. Then, in the $Q\bar{Q}$ center of mass of mass frame one can write the multichannel Schr\"{o}dinger equation
\begin{equation}
\label{SEP2}
(\mathrm{K}+\mathrm{V})\ket{\psi}=E\ket{\psi}
\end{equation}
where $\ket{\psi}$ is the heavy-quark meson state
\[
\ket{\psi}=
\begin{pmatrix}
\ket{\psi^{(0)}}\\
\ket{\psi^{(1)}}\\
\ket{\psi^{(2)}}
\end{pmatrix}
\]
with the superscript $0$ and $i$ referring, respectively, to its $Q\bar{Q}$ and $M^{(i)}\bar{M}^{(i)}$ component, $\mathrm{K}$ the relative kinetic energy, $\mathrm{V}$ the (diabatic) potential, and $E$ the energy.

In order to do a systematic study of \eqref{SEP2}, we proceed to an analysis of the Hamiltonian in terms of powers of the inverse of the heavy quark mass, $1 / m_{Q}$. In this manner, the relative relevance of the different interactions in the Hamiltonian can be established. This is of interest to identify the dominant dynamical mechanisms not only in this case, but also if additional components and interactions were considered.

So, for the relative kinetic energy, we have
\[
\mathrm{K}=
\begin{pmatrix}
\frac{p^{2}}{m_{Q}} & 0 & 0\\
0 & \frac{p^{2}}{2\mu^{(1)}} & 0\\
0 & 0 & \frac{p^{2}}{2\mu^{(2)}}
\end{pmatrix}
\]
with $\bm{p}$ standing for the relative momentum operator and $\mu$ for the reduced mass. Notice that, up to order $1 / m_{Q}$, one has $1 / (2\mu^{(i)})\simeq1 / m_{Q}$ and $\mathrm{K}\simeq \mathbbold{1} {p^{2}} / m_{Q}$, with $\mathbbold{1}$ the identity matrix.

 %\bigskip

The potential $\mathrm{V}$ results from integrating out the light degrees of freedom. This Hermitian matrix is the representation of $H_\textup{static}$, the light-field Hamiltonian containing also the interaction with the $Q\bar{Q}$ and $M^{(i)}\bar{M}^{(i)}$ components, in the basis $\{\ket{ \zeta_{Q\bar{Q}}}\} \cup \{\ket{\zeta_{M^{(i)}\bar{M}^{(i)}}}\}_{i=1,2}$, formed by the light field eigenstates in the absence of any mixing interaction \cite{Bru20}. It reads
\[
\mathrm{V}=
\begin{pmatrix}
V_{00} & V_{01} & V_{02}\\
V_{01}^{\dag} & V_{11} & 0\\
V_{02}^{\dag} & 0 & V_{22}
\end{pmatrix}
\]
with
\begin{align*}
V_{00}&=\braket{\zeta_{Q\bar{Q}}\rvert H_\textup{static} \lvert\zeta_{Q\bar{Q}}}, \\
V_{ii}&=\braket{\zeta_{M^{(i)}\bar{M}^{(i)}}\rvert H_\textup{static} \lvert\zeta_{M^{(i)}\bar{M}^{(i)}}}, \\
V_{0i}&=\braket{\zeta_{Q\bar{Q}} \rvert H_\textup{static} \lvert\zeta_{M^{(i)}\bar{M}^{(i)}}}.
\end{align*}
Let us note that we have put $V_{12}=0$, meaning that no direct mixing between different meson-meson components through $H_\textup{static}$ is considered, and consequently no direct coupling of $\ket{\psi^{(1)}}$ with $\ket{\psi^{(2)}}$ is present.

A key point of this diabatic formalism is that the functional form of elements of the diabatic potential matrix $\mathrm{V}$ can be obtained from currently available Lattice results. Specifically, the diagonal element $V_{00}$ has been calculated \textit{ab initio} in quenched (no sea quarks) Lattice QCD, as the energy of the ground state of the gluon field in the presence of $Q$ and $\bar{Q}$ color sources placed at a relative distance $r$. Up to order $1 / m_{Q}$ (for the specific form of the $1 / m_{Q}^{2}$ spin dependent terms, see \cite{Eic81}), it corresponds to a central potential whose form mimics the phenomenological Cornell potential \cite{Bal01}
\[
V_\textup{C}(r)=\sigma r-\frac{\chi}{r} + 2 m_{Q} - \beta.
\]
It is worth recalling that in phenomenological applications of this potential, as the calculation of the spectrum of conventional quarkonium, the string tension $\sigma$, color Coulomb strength $\chi$, and heavy quark mass $m_{Q}$ are effective parameters whose values may be incorporating some effect from the terms of order $1 / m_{Q}^{2}$ and higher. The values of $\sigma$ and $\chi$ are usually assumed to be flavor independent, and this can also be the case for the constant $\beta$ through a convenient choice of the values of the quark masses.

 %\bigskip

For the other diagonal elements $V_{ii}$ we shall follow, for the sake of simplicity, a free meson-meson approximation, implying
\[
V_{ii} = T^{(i)}
\]
where $T^{(i)}= m_{M^{(i)}}+m_{\bar{M}^{(i)}} = 2 m_{Q} + c^{(i)} + \bar{c}^{(i)}$ is the meson-meson threshold. To first order in $1 / m_{Q}$, it can be expressed as $T^{(i)}\simeq 2m_{Q}+a+{b} / m_{Q}$ where $a$ does not depend on $m_{Q}$ and $b$ has at most some logarithmic dependence \cite{Bra18b}.

 %\bigskip

As for the off-diagonal elements $V_{0i}$, their radial dependence can be derived from the results of unquenched (with sea quarks) Lattice calculations. More concretely, the energy levels of the light fields in the presence of $Q$ and $\bar{Q}$ color sources placed at a relative distance $r$, when $\ket{\zeta_{Q\bar{Q}}}$ and $\ket{\zeta_{M^{(i)}\bar{M}^{(i)}}}$ mix, have been calculated \cite{Bal05,Bul19}. To connect these Lattice results with $V_{0i}$, let us particularize, for reasons that shall be made clear later on, to the case $Q\equiv b$, $M^{(1)}\bar{M}^{(1)} \equiv B^{+} B^{-}$, and $M^{(2)}\bar{M}^{(2)} \equiv B^{0} \bar{B}^{0}$. Let us also realize that the $B^{+}B^{-}$ and $B^{0}\bar{B}^{0}$ thresholds are approximately degenerate, $T_{B^{+}B^{-}}\approx T_{B^{0}\bar{B}^{0}} = T_{B\bar{B}}$ with $T_{B\bar{B}}$ a shorthand notation for the degenerate threshold mass. Then, we can build light-field states with definite isospin, ($I=0,I_{z}=0$) and ($I=1,I_{z}=0$):
\begin{align*}
\ket{\zeta_{(B\bar{B})_{I=0}}} &= \frac{1}{\sqrt{2}} \bigl( \ket{\zeta_{B^{+}B^{-}}} + \ket{\zeta_{B^{0}\bar{B}^{0-}}} \bigr), \\
\ket{\zeta_{(B\bar{B})_{I=1}}} &= \frac{1}{\sqrt{2}} \bigl( \ket{\zeta_{B^{+}B^{-}}} - \ket{\zeta_{B^{0}\bar{B}^{0-}}} \bigr),
\end{align*}
so that $\braket{\zeta_{b\bar{b}} \rvert H_\textup{static} \lvert \zeta_{(B\bar{B})_{I=0}}}= (V_{01}+V_{02}) / \sqrt{2}$ and $\braket{\zeta_{b\bar{b}} \rvert H_\textup{static} \lvert \zeta_{(B\bar{B})_{I=1}}}=(V_{01}-V_{02}) / \sqrt{2}$.

As $b\bar{b}$ has $I=0$ and $H_\textup{static}$ does not contain any isospin breaking term (once the thresholds of $B^{+}B^{-}$ and $B^{0}\bar{B}^{0}$ are taken to be equal), $\ket{\zeta_{(B\bar{B})_{I=1}}}$ cannot couple to $b\bar{b}$. This implies $V_{01}=V_{02}$, so that, in the isospin basis, $\mathrm{V}$ can be written as
\[
\begin{pmatrix}
V_\textup{C} & \sqrt{2}V_{01} & 0\\
\sqrt{2}V_{01}^{\dag} & T_{B\bar{B}}
& 0\\
0 & 0 & T_{B\bar{B}}
\end{pmatrix}.
\]
This makes clear that the effect of the degenerate $B^{+}B^{-}$ and $B^{0}\bar{B}^{0}$ thresholds can be taken into account through only one isospin-zero threshold, that we shall call henceforth $B\bar{B}$, whose interaction potential with $b\bar{b}$ contains an additional factor $\sqrt{2}$ as compared to that of a non-degenerate threshold like $B_{s}\bar{B}_{s}$.

\begin{figure}
\centering
\includegraphics{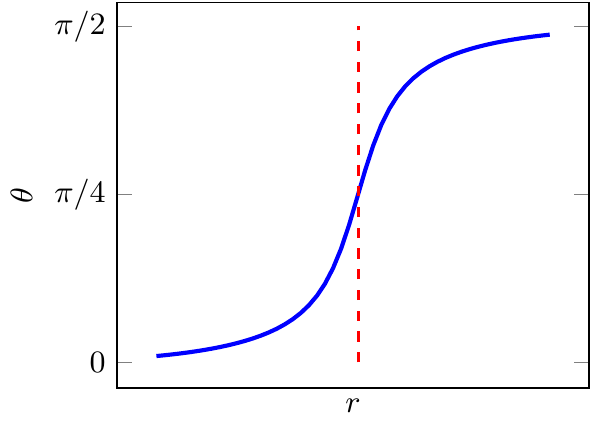}
\caption{\label{angle}Mixing angle $\theta$, in radians, as a function of the distance $r$. The crossing radius is highlighted by the dashed vertical line.}
\end{figure}

\begin{figure}
\centering
\includegraphics{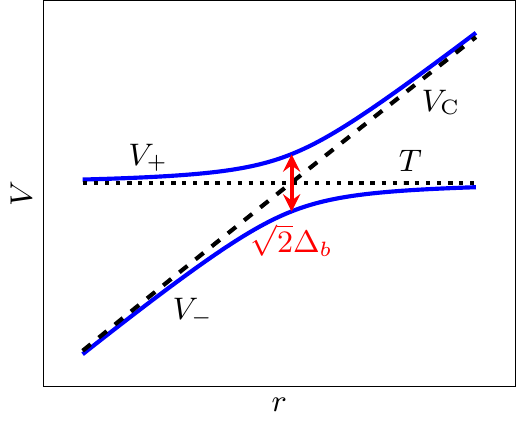}
\caption{\label{cross}Schematic representation of the static energies near the avoided crossing. Solid lines: unquenched ground and excited state light-field energies. Dashed line: quenched ground state static light-field energy. Dotted line: meson-meson threshold.}
\end{figure}

Actually, in \cite{Bal05} the radial dependence of the eigenstates and eigenvalues of
\[
\begin{pmatrix}
V_\textup{C} & \sqrt{2}V_{01}\\
\sqrt{2}V_{01}^{\dag} & V_{B\bar{B}}
\end{pmatrix},
\]
where $V_{B\bar{B}}$ differs from $T_{B\bar{B}}$ in the incorporation of the meson-meson interaction, has been calculated. For the eigenstates, it can be expressed as
\begin{subequations}
\label{Mix}
\begin{align}
\ket{\zeta_{-}} &= \cos\theta(r)\ket{\zeta_{b\bar{b}}}+\sin\theta(r)\ket{\zeta_{B\bar{B}}} \\
\ket{\zeta_{+}} &= -\sin\theta(r)\ket{\zeta_{b\bar{b}}}+\cos\theta(r)\ket{\zeta_{B\bar{B}}}
\end{align}
\end{subequations}
where $\theta(r)$ is the mixing angle. Following \cite{Bal05}, we shall consider that the effect of the meson-meson interaction, $V_{B\bar{B}} - T_{B\bar{B}}$, in $\theta(r)$ is the appearance of a bump for low values of $r$ (in \cite{Bul19} no bump appears). A schematic representation of $\theta(r)$ without the bump is presented in Fig.~\ref{angle}. A look at this figure makes clear that the maximum or ideal ($\theta=\pi / 4$) light field configuration mixing occurs at approximately the crossing radius $r_{B\bar{B}}$, defined by $V_\textup{C}(r_{B\bar{B}})=T_{B\bar{B}}$, and that the mixing is only significant around $r_{B\bar{B}}$. The corresponding eigenvalues, $V_{-}$ and $V_{+}$, are schematically represented in Fig.~\ref{cross}. The minimal energy gap between them, $\sqrt{2} \Delta_{b}$, occurring at $r_{B\bar{B}}$ is also drawn.

From $\braket{\zeta_{\pm} \rvert H_\textup{static} \rvert \zeta_{\pm}}=V_{\pm}$, $\braket{\zeta_{\pm} \rvert H_\textup{static} \rvert \zeta_{\mp}}= 0$, and Eq.~\eqref{Mix}, the radial dependence of the mixing potential $V_{01}$, that we shall call $V_{B\bar{B}}^\textup{mix}(r)$, can be straightforwardly extracted (for the complete spatial dependence of $V_{01}$, see \cite{Bru21c}). It reads
\[
V_{B\bar{B}}^\textup{mix}(r)= - \frac{\sin2\theta(r)}{2\sqrt{2}}(V_{+}(r)-V_{-}(r)),
\]
so that its absolute value is significant only around $r_{B\bar{B}}$, where it reaches a maximum $\lvert V_{B\bar{B}}^\textup{mix}(r_{B\bar{B}}) \rvert=\Delta_{b}/2$ with $\Delta_{b}= (V_{+}(r_{B\bar{B}})-V_{-}(r_{B\bar{B}})) / \sqrt{2}$.

The term $\sqrt{2}V_{B\bar{B}}^\textup{mix}(r)$ determines the mixing between the $b\bar{b}$ and $B\bar{B}$ components. The physical mechanism underlying this mixing is string breaking. As the static energy of $b\bar{b}$, $V_\textup{C}(r)$, approaches that of $B\bar{B}$, $T_{B\bar{B}}$, an interaction between the two components, from the creation of a light quark pair $q\bar{q}$ and its recombination with $b\bar{b}$, becomes more and more probable. This makes, see Fig.~\ref{cross}, that the static energies, which in the absence of string breaking would cross each other at $r_{B\bar{B}}$, experience an avoided crossing characterized by a nonvanishing energy gap $\sqrt{2}\Delta_{b}= V_{+}(r_{B\bar{B}})-V_{-}(r_{B\bar{B}})\neq0$.

In general, string breaking is expected to occur as a consequence of the $Q\bar{Q}$--meson-meson threshold interaction. According to our discussion above, we expect $V_{M^{(i)}\bar{M}^{(i)}}^\textup{mix}(r)$ to be a function of $V_\textup{C}(r)-T^{(i)}$. Besides, Lattice results for $\theta(r)$ and $V_{+}(r)-V_{-}(r)$ are approximately symmetric with respect to the crossing radius (see Figs. 16 and 17 in \cite{Bal05}). Then, the simplest parametrization may be
\[
V_{M^{(i)}\bar{M}^{(i)}}^\textup{mix}(r) \simeq -\frac{\Delta_{Q}}{2}f(V_\textup{C}(r)-T^{(i)}),
\]
where $f$ is a positive even function with an absolute maximum, $f(0) =1$, that vanishes for $V_\textup{C}(r) \ll T^{(i)}$ and $V_\textup{C}(r) \gg T^{(i)}$, and we have assumed $\Delta_{Q}$ to be the same for any of the corresponding thresholds.

Notice that in the limit $m_{Q}\to\infty$, which implies $m_{M^{(i)}}\to\infty$, a single-channel (i.e.: no mixing) approximation must be recovered since, as shown in \cite{Bru20}, the nonadiabatic coupling terms breaking it are weighted by a factor $1/m_{Q}$, see Eq.~(17) in \cite{Bru20}. Hence, $\lim_{m_{Q}\to\infty}\Delta_{Q}=0$ and we can expand $\Delta_{Q}$ in powers of $1 / m_{Q}$ as $\Delta_{Q}= \alpha / m_{Q} + \mathcal{O}(1 / m_{Q}^{2})$, where $\alpha$ is a constant with dimensions of energy squared. Then, at order $1 / m_{Q}$, we predict
\begin{equation}
\label{result1}
\Delta_{c}m_{c}\simeq\Delta_{b}m_{b}.
\end{equation}

Let us also realize that $\alpha$ has the same dimensions as $\sigma$, and that if there were no confining interaction, no avoided crossing could ever take place. Therefore, it is natural to express $\alpha$ as $\sigma$ times a dimensionless constant $\gamma$, so that, up to order $1 / m_{Q}$, we have
\begin{equation}
\label{result2}
\Delta_{Q}\simeq\frac{\gamma \sigma}{m_{Q}}
\end{equation}
and
\begin{equation}
\label{result3}
V_{M^{(i)}\bar{M}^{(i)}}^\textup{mix}(r)\simeq-\frac{\gamma \sigma}{2 m_{Q}}f(V_\textup{C}(r)-T_{M^{(i)}\bar{M}^{(i)}}).
\end{equation}

Equations \eqref{result1}, \eqref{result2}, and \eqref{result3}, where $\gamma\approx1$ as we show next, represent a main outcome of this letter. They tell us that in a systematic expansion of the Hamiltonian of a heavy-quark meson in powers of $1 / m_{Q}$, the dominant order in the string breaking mixing potential is $1 / m_{Q}$ with a strength proportional to the string tension.

In practical applications, as previously mentioned, $\sigma$, $\chi$, $\beta$ and $m_{Q}$ are effective parameters which are fine tuned phenomenologically.
As for $\gamma$, it may be fixed by matching $\Delta_{Q}$ with the energy gap $\Delta_\textup{Lat}$ calculated in Lattice QCD, at some finite heavy quark mass that we call $\widetilde{m}_{Q}$. As a matter of fact, using standard phenomenological values \cite{Eic94} $\sigma=(427.4\text{ MeV})^{2}$ and $\chi=0.52$, a slightly modified quark mass $m_{b}=5215$~MeV to get the same $\beta=855$~MeV for charmonium and bottomonium, and $\sqrt{2}\Delta_{b} = 51~\text{MeV} \approx \Delta_\textup{Lat}$ from \cite{Bal05}, a nice (diabatic) description of the spectrum and properties of bottomoniumlike mesons comes out \cite{Bru21b} (we have checked that the slight difference in the value of $\Delta_{b}$ used in \cite{Bru21b} with respect to $\sqrt{2}\Delta_{b}=51$~MeV gives rise to spectral mass differences of $2$~MeV at most). Hence, we may identify $\widetilde{m}_{Q} \approx m_{b}$, so that using \eqref{result2} we get $\gamma \approx 1.03$.

Then, taking into account that for the same standard values of $\sigma$, $\chi$ and $\beta$, the charm quark mass is quite constrained from electromagnetic transitions in charmonium \cite{Bru19b} to be $m_{c}\approx 1840$~MeV, we predict from \eqref{result1} $\Delta_{c}\approx102.2$~MeV.

In order to check whether this value of $\Delta_{c}$ may give rise to $\chi_{c1}(3872)$, we proceed to solve \eqref{SEP2}, expanding the Hamiltonian up to order $1 / m_{Q}$, to get the $J^{PC}=1^{++}$ bound states in exactly the same manner as done in \cite{Bru20}, to which we refer for details. There are, however, two crucial differences. First, the value of $\Delta_{c}$ is now a theoretical input, instead of a parameter fitted to get the $\chi_{c1}(3872)$. Second, the different masses of the $D^{0}\bar{D}^{\ast 0}$ and $D^{+}D^{\ast-}$ thresholds are implemented, instead of an effective isospin-zero $D\bar{D}^{\ast}$ component, where $D^{0}\bar{D}^{\ast 0}$, $D^{+}D^{\ast-}$, and $D\bar{D}^{\ast}$, is the common shorthand notation for the $C$-parity eigenstates.

For the sake of completeness, we also list the values of the meson masses: $m_{D^{0}}=1864.8$~MeV, $m_{\bar{D}^{\ast 0}}=2006.9$~MeV, $m_{D^{+}}=1869.5$~MeV, $m_{D^{\ast-}}=2010.3$~MeV. It is worth to comment that that the use of these open-charm meson experimental masses has implicit the assumption that terms of order $1 / m_{c}^{2}$ and higher are negligible for them. As for the radial mixing potential, we assume the Gaussian form
\[
V_{M^{(i)}\bar{M}^{(i)}}^\textup{mix}(r)=-\frac{\Delta_{c}}{2}\exp\biggl\{-\frac{(V_\textup{C}(r)-T^{(i)})^{2}}{2\Lambda^{2}}\biggr\}
\]
where $\Lambda = \sigma\rho$ with $\rho \approx 0.3$~fm the radial scale for the mixing, which we assume to be the same for $Q\equiv c$ and $Q\equiv b$. Notice that from this Gaussian form the corrections of order $1 / m_{c}^{2}$ and higher in $(V_\textup{C}(r) - T_{M^{(i)}\bar{M}^{(i)}})^{2}$,
which cannot be easily separated in the calculation, hardly play any quantitative role.

\begin{figure}
\centering
\includegraphics{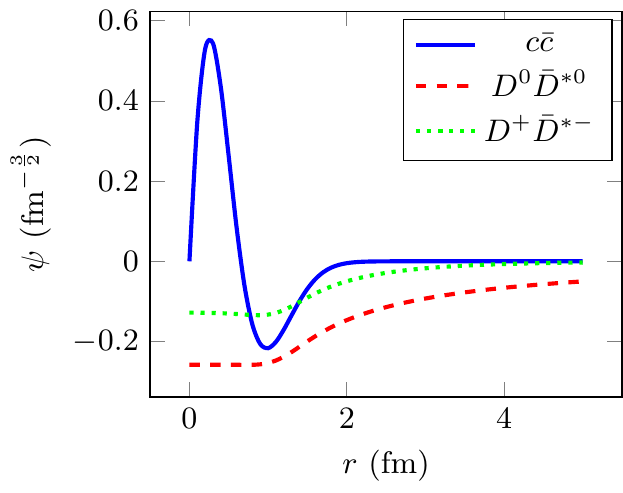}
\caption{\label{x3872}Diabatic radial wave function of $\chi_{c1}(3872)$.}
\end{figure}

We find a $J^{PC}=1^{++}$ bound state at $3871.6$~MeV, which can be assigned to $\chi_{c1}(3872)$. The calculated probabilities for the different components, $P_{c\bar{c}}=4\%$, $P_{D^{0}\bar{D}^{\ast0}}=93\%$, and $P_{D^{+}D^{\ast-}}=3\%$, indicate that $\chi_{c1}(3872)$ is very dominantly a $D^{0}\bar{D}^{\ast0}$ state. However, it should be pointed out that at short distances, $r \lesssim 1$~fm, the $c\bar{c}$ and $D^{0}\bar{D}^{\ast0}$ probability densities are comparable, as shown in Fig.~\ref{x3872}, what may be instrumental to explain the electromagnetic decays of $\chi_{c1}(3872)$. In contrast, at distances $r \gtrsim 3$~fm, only the $D^{0}\bar{D}^{\ast0}$ component survives. The large value of the root mean square radius, $\sqrt{\langle r^{2} \rangle} \approx 14$~fm, supports the physical image of $\chi_{c1}(3872)$ as a loosely bound $D^{0}\bar{D}^{\ast0}$ molecule. Moreover, the equal content of isospin 0 and isospin 1 in $D^{0}\bar{D}^{\ast0}$ may provide us with a natural explanation for the approximately equal measured decay rates of $\chi_{c1}(3872)$ to $\omega J\!/\!\psi$ and $\pi^{+}\pi^{-} J\!/\!\psi$. Let us emphasize that this state is a prediction, not a fit, as the value of $\Delta_{c}$ is completely fixed from the matching of $\Delta_{b}$ with $\Delta_\textup{Lat}$.

It is nevertheless important to study the robustness of this prediction against small variations of $\Delta_{c}$ around its nominal value. A numerical analysis shows that for $\Delta_{c} \gtrsim 101$~MeV the bound state prediction is stable, its binding energy increasing with $\Delta_{c}$. Moreover, for $\Delta_{c}\in[101,103]$~MeV, the predicted bound state mass is perfectly compatible with the PDG average experimental mass of $\chi_{c1}(3872)$, $3871.65\pm0.06$~MeV.

The resulting physical image of $\chi_{c1}(3872)$ is in line with the one from molecular models, where hadronic molecules are the result of nonperturbative meson-meson interactions (see, for example, \cite{Hosaka16,Dong17,Guo18} and references therein). A more quantitative connection with these models can be done by realizing that, as explicitly shown in a recent study of meson-meson scattering in the diabatic framework \cite{Bru21c}, the diabatic potential gives rise to a nonperturbative meson-meson interaction mediated by $Q\bar{Q}$ (see Fig.~1b in Ref.~\cite{Bru21c}). Concretely, we have repeated the calculation carried out in \cite{Bru21c}, but considering $D^{0}\bar{D}^{\ast0}$ and $D^{+}D^{\ast-}$ as separate channels with nondegenerate thresholds, instead of an effective $D\bar{D}^{\ast}$ channel with isospin zero. Although this analysis is completely out of the scope of this letter, and will be included in a future publication, let us just advance that for $\Delta_{c}\gtrsim 101$~MeV the calculated S matrix consistently reflects the presence of the bound state close below threshold, while for $\Delta_{c}\lesssim 101$~MeV we get instead a virtual state.

It is also important to emphasize that, although we restrict here our analysis to the $\chi_{c1}(3872)$, a consistent description of the whole charmoniumlike spectrum, with no significant difference with respect to the one obtained in \cite{Bru20}, comes out. Furthermore, systematic corrections could be incorporated through higher order terms in the Hamiltonian expansion.

 %\bigskip

Therefore, we conclude that $\chi_{c1}(3872)$ (as well as other unconventional heavy-quark mesons) may be generated from the mixing of $Q\bar{Q}$ with open-flavor meson-meson components induced by string breaking.

\begin{acknowledgments}
This work has been supported by \foreignlanguage{spanish}{Ministerio de Ciencia e Innovaci\'on} and \foreignlanguage{spanish}{Agencia Estatal de Investigaci\'on} of Spain MCIN/AEI/10.13039/501100011033 and European Regional Development Fund Grant No.~PID2019-105439~GB-C21, by EU Horizon 2020 Grant No.~824093 (STRONG-2020), and by \foreignlanguage{spanish}{Conselleria de Innovaci\'on, Universidades, Ciencia y Sociedad Digital, Generalitat Valenciana} GVA~PROMETEO/2021/083. R.B. acknowledges a FPI fellowship from \foreignlanguage{spanish}{Ministerio de Ciencia, Innovaci\'on y Universidades} of Spain under Grant No.~BES-2017-079860.
\end{acknowledgments}

\bibliography{mixbib}

\end{document}